\documentclass[aps,pra,twocolumn,superscriptaddress,showpacs]{revtex4}
\usepackage{amssymb}
\usepackage{amsmath}
\usepackage{graphicx}
\usepackage{color}

\begin{document}

\title{Influence of the lighting on Fast Atom Diffraction studied via a
semi-quantum approach}
\author{M.S. Gravielle\thanks{%
Author to whom correspondence should be addressed.\newline
Electronic address: msilvia@iafe.uba.ar}}
\affiliation{Instituto de Astronom\'{\i}a y F\'{\i}sica del Espacio (IAFE, CONICET-UBA),
casilla de correo 67, sucursal 28, C1428EGA, Buenos Aires, Argentina.}
\author{J.E. Miraglia}
\affiliation{Instituto de Astronom\'{\i}a y F\'{\i}sica del Espacio
(IAFE, CONICET-UBA), casilla de correo 67, sucursal 28, C1428EGA,
Buenos Aires, Argentina.}
\date{\today }

\begin{abstract}
The influence of the collimating conditions of the incident beam on
diffraction patterns produced by grazing scattering of fast atoms off
crystal surfaces is studied within a semi-quantum approach, named Surface
Initial Value Representation (SIVR) approximation. In this approach we
incorporate a realistic description of the incident particle in terms of the
collimating parameters, which determine the surface area that is coherently
illuminated. The model is applied to He atoms colliding with a LiF(001)
surface after passing through a rectangular aperture. As it was
experimentally observed \cite{Winter15}, SIVR spectra as a function of the
azimuthal angle are very sensitive to the width of the collimating slit. We
also found that the length of the collimating aperture affects \ polar angle
distributions, introducing additional interference structures for the longer
collimating slits.
\end{abstract}

\pacs{34.35.+a,79.20.Rf, 37.25.+k } \maketitle

\section{Introduction}

Diffraction patterns produced by grazing scattering of swift atoms and
molecules (with energies in the keV range) on surfaces are nowadays becoming
a powerful surface analysis tool, which is giving rise to a technique known
as grazing-incidence fast atom diffraction (GIFAD or FAD) \cite%
{Winter11,Zugarramurdi15}. In recent years the FAD method was successfully
applied to very different kinds of materials, ranging from insulators \cite%
{Schuller07,Rousseau07,Schuller12} to semiconductors \cite%
{Khemliche09,Debiossac14} and metals \cite{Bundaleski08,Busch09,Rubiano13},
as well as structured \ films \cite{Seifert10} and molecules \cite%
{SeifertPRL13} adsorbed on surfaces. However, in spite of the extensive
experimental and theoretical work devoted to the research of FAD since its
first experimental observation \cite{Schuller07,Rousseau07}, the complete
understanding of the underlying quantum processes is far from being
achieved. In particular, the study of the mechanisms that contribute to the
coherence or decoherence of the scattered particles is\ still in its infancy.

The observation of quantum interference effects for fast atoms impinging on
crystal surfaces strongly relies on the preservation of quantum coherence
\cite{Aigner08,Lienemann11,Bundaleski11} and in this regard, the coherence
conditions of the incident beam play an important role. Motivated by Ref.
\cite{Winter15}, in this article we investigate the influence of the
collimation of the incident beam on FAD patterns making use of a recently
developed approach, named Surface-Initial Value Representation (SIVR)
approximation \cite{Gravielle14}. With this goal we explicitly take into
account the experimental collimating conditions to determine the surface
region that is \textit{coherently }illuminated by the particle beam, using
this information to build the initial wave packet that describes the
unperturbed state of the incident particle within the SIVR method.

The SIVR approximation is a semi-quantum approach that was derived from the
Initial Value Representation (IVR) method by Miller \cite{Miller70} by using
the corresponding semi-quantum time evolution operator in the frame of a
time-dependent distorted-wave formalism. This strategy incorporates an
approximate description of classically forbidden transitions on the dark
side of rainbow angles, making it possible to avoid the classical rainbow
divergence present in previous semi-classical models for FAD, like the
Surface-Eikonal (SE) approach \cite{Gravielle08,SchullerGrav09}. Such a
weakness of the SE method affects the intensity of the outermost diffraction
maxima when these maxima are close to the classical rainbow angles \cite%
{Rubiano13}, i.e. the extreme deflection angles of the classical projectile
distribution. The SIVR approach, instead, provides an appropriate
description of FAD patterns along the whole angular range, even around
classical rainbow angles, without requiring the use of convolutions to
smooth the theoretical curves \cite{Gravielle14}. Therefore, the SIVR method
can be considered as an attractive alternative to quantum wave packet
propagations, offering a clear representation of the main mechanisms of the
process in terms of classical trajectories through the Feynman path integral
formulation of quantum mechanics.

In order to analyze the influence of the beam collimation on FAD spectra, an
extended version of the SIVR approximation - including the collimating
parameters - is applied to evaluate FAD patterns for He atoms grazingly
impinging on a LiF(001) surface after going through a rectangular aperture.
The paper is organized as follows. The theoretical formalism is summarized
in Sec. II. Results for different sizes of the collimating aperture are
presented and discussed in Sec. III, while in Sec. IV we outline our
conclusions. Atomic units (a.u.) are used unless otherwise stated.

\section{Theoretical model}

Let us consider an atomic projectile ($P$), with initial momentum $\vec{K}%
_{i}$, which is elastically scattered from a crystal surface ($S$), ending
in a final state with momentum $\vec{K}_{f}$ \ and total energy $%
E=K_{f}^{2}/(2m_{P})=K_{i}^{2}/(2m_{P})$, with $m_{P}$ the projectile mass.
By employing the IVR method \cite{Miller01}, the scattering state of the
projectile at the time $t$ \ can be approximated as \cite{Gravielle14}:

\begin{eqnarray}
\left\vert \Psi _{i}^{{\small (SIVR)}+}(t)\right\rangle &=&\frac{1}{(2\pi
i)^{3/2}}\int d\overrightarrow{R}_{o}\ f_{i}(\overrightarrow{R}_{o})\int d%
\overrightarrow{K}_{o}\ g_{i}(\overrightarrow{K}_{o})  \notag \\
&&\times \ \left( J_{{\small M}}(t)\right) ^{1/2}\ \Phi _{i}(\overrightarrow{%
R}_{o})\exp (iS_{t})\left\vert \mathcal{\vec{R}}_{t}\right\rangle ,  \notag
\\
&&  \label{estado-ivr}
\end{eqnarray}%
where
\begin{equation}
\Phi _{i}(\vec{R})=(2\pi )^{-3/2}\exp (i\vec{K}_{i}\cdot \vec{R})\quad
\label{fi-i}
\end{equation}%
is the initial momentum eigenfunction, with $\vec{R}\ $ the position of the
center of mass of the incident atom, and the sign "$+$" in the supra-index
of the scattering state indicates that it satisfies outgoing asymptotic
conditions. In Eq. (\ref{estado-ivr}) the position ket $\left\vert \mathcal{%
\vec{R}}_{t}\right\rangle $ is associated with the time-evolved position of
the incident atom at a given time $t$, $\mathcal{\vec{R}}_{t}\equiv \mathcal{%
\vec{R}}_{t}(\overrightarrow{R}_{o},\overrightarrow{K}_{o})$, which is
derived by considering a classical trajectory with starting position and
momentum $\overrightarrow{R}_{o}$ and $\overrightarrow{K}_{o}$,
respectively. The function $S_{t}$ denotes the classical action along the
trajectory
\begin{equation}
S_{t}=S_{t}(\overrightarrow{R}_{o},\overrightarrow{K}_{o})=\int%
\limits_{0}^{t}dt^{\prime }\ \left[ \frac{\overrightarrow{\mathcal{P}}%
_{t^{\prime }}^{2}}{2m_{P}}-V_{SP}(\mathcal{\vec{R}}_{t^{\prime }})\right] ,
\label{St}
\end{equation}%
with $\overrightarrow{\mathcal{P}}_{t}=m_{P}d\mathcal{\vec{R}}_{t}/dt\ $ the
classical projectile momentum at the time $t$ and $V_{SP}$ the
surface-projectile interaction, while the function
\begin{equation}
J_{{\small M}}(t)=\det \left[ \frac{\partial \mathcal{\vec{R}}_{t}(%
\overrightarrow{R}_{o},\overrightarrow{K}_{o})}{\partial \overrightarrow{K}%
_{o}}\right]  \label{J}
\end{equation}%
is a Jacobian factor (a determinant) evaluated along the classical
trajectory $\mathcal{\vec{R}}_{t}$. This Jacobian factor can be related to
the Maslov index \cite{Guantes04} by expressing it as $J_{{\small M}%
}(t)=\left\vert J_{M}(t)\right\vert \exp (i\nu _{t}\pi )$, where $\left\vert
J_{{\small M}}(t)\right\vert $ is the modulus of $J_{M}(t)$ and $\nu _{t}$
is an integer number that accounts for the sign of $J_{M}(t)$ at a given
time $t$. In this way, $\nu _{t}$ represents a time-dependent Maslov index,
satisfying that every time that $J_{{\small M}}(t)$ changes its sign along
the trajectory, $\nu _{t}$ increases by 1.

The functions $f_{i}(\overrightarrow{R}_{o})$ and $g_{i}(\overrightarrow{K}%
_{o})$, present in the integrand of Eq. (\ref{estado-ivr}), describe the
shape of the position- and momentum- wave packet associated with the
incident projectile. In a previous paper \cite{Gravielle14} $f_{i}(%
\overrightarrow{R}_{o})$ was considered as a Gaussian distribution
illuminating a fixed number of reduced unit cells of the crystal surface,
while $g_{i}(\overrightarrow{K}_{o})$ was defined as an uniform
distribution. Here these functions are derived from the collimation
conditions of the incident beam in order to \ incorporate a realistic
profile of the coherent initial wave packet, as explained in the following
sub-section.

By using the SIVR scattering state, given by Eq. \ (\ref{estado-ivr}),
within the framework of the time-dependent distorted-wave formalism \cite%
{Dewangan94}, the SIVR transition amplitude,\ per unit of surface area $%
\mathcal{S}$, can be expressed as \cite{Gravielle14}
\begin{eqnarray}
A_{if}^{{\small (SIVR)}} &=&\frac{1}{\mathcal{S}}\int\limits_{\mathcal{S}}d%
\overrightarrow{R}_{o}\ f_{i}(\overrightarrow{R}_{o})\int d\overrightarrow{K}%
_{o}\ g_{i}(\overrightarrow{K}_{o})  \notag \\
&&\times \ a_{if}^{{\small (SIVR)}}(\overrightarrow{R}_{o},\overrightarrow{K}%
_{o}),  \label{Aif-sivr}
\end{eqnarray}%
where
\begin{eqnarray}
a_{if}^{{\small (SIVR)}}(\overrightarrow{R}_{o},\overrightarrow{K}_{o}) &=&\
-\int\limits_{0}^{+\infty }dt\ \ \frac{\left\vert J_{M}(t)\right\vert
^{1/2}e^{i\nu _{t}\pi /2}}{(2\pi i)^{9/2}}V_{SP}(\mathcal{\vec{R}}_{t})
\notag \\
&&\times \exp \left[ i\left( \varphi _{t}^{{\small (SIVR)}}-\overrightarrow{Q%
}\cdot \overrightarrow{R}_{o}\right) \right] \quad  \label{aif}
\end{eqnarray}%
is the partial transition amplitude associated with the classical path $%
\mathcal{\vec{R}}_{t}\equiv \mathcal{\vec{R}}_{t}(\overrightarrow{R}_{o},%
\overrightarrow{K}_{o})$, with $\overrightarrow{Q}=\vec{K}_{f}-\vec{K}_{i}$
the projectile momentum transfer and
\begin{equation}
\varphi _{t}^{{\small (SIVR)}}=\int\limits_{0}^{t}dt^{\prime }\ \left[ \frac{%
1}{2m_{P}}\left( \vec{K}_{f}-\overrightarrow{\mathcal{P}}_{t^{\prime
}}\right) ^{2}-V_{SP}(\mathcal{\vec{R}}_{t^{\prime }})\right]  \label{fitot}
\end{equation}%
the SIVR phase at the time $t$. Details of the derivation of the SIVR method
are given in Ref. \cite{Gravielle14}.

In this article we use a frame of reference placed on the first atomic
layer, with the surface contained in the $x-y$ plane, the $\widehat{x}$
versor along the incidence direction and the $\hat{z}$ versor oriented
perpendicular to the surface, aiming towards the vacuum region. The SIVR
differential probability, per unit of surface area, for elastic scattering
with final momentum $\vec{K}_{f}$ in the direction of the solid angle $%
\Omega _{f}\equiv (\theta _{f},\varphi _{f})$, is obtained from Eq. (\ref%
{Aif-sivr}) as
\begin{equation}
dP^{{\small (SIVR)}}/d\Omega _{f}=K_{f}^{2}\left\vert A_{if}^{{\small (SIVR)}%
}\right\vert ^{2},  \label{dPdangle}
\end{equation}%
where $\theta _{f}$ and $\varphi _{f}$ are the final polar and azimuthal
angles, respectively, with $\theta _{f}$ measured with respect to the
surface and $\varphi _{f}$ \ measured with respect to the $\widehat{x}$
axis. A schematic depiction of the process and the coordinates is displayed
in Fig. 1 (a).

\subsection{Size of the \textit{coherent} initial wave packet}

In Eq. (\ref{Aif-sivr}), the variables $\overrightarrow{R}_{o}$ and $%
\overrightarrow{K}_{o}$ represent the starting position and momentum,
respectively, of the classical projectile trajectory, both measured at $t=0$%
, while the functions $f_{i}(\overrightarrow{R}_{o})$ and $g_{i}(%
\overrightarrow{K}_{o})$ determine the shape of the initial wave packet,
satisfying the Heisenberg uncertainty relation. We discompose the starting
position as $\ \overrightarrow{R}_{o}=\overrightarrow{R}_{os}+Z_{o}\widehat{z%
}$, where $\overrightarrow{R}_{os}=$ $X_{o}\widehat{x}+Y_{o}\widehat{y}$ and
$Z_{o}$ are the components parallel and perpendicular, respectively, to the
surface plane, with $Z_{o}$ a fixed distance for which the projectile is
hardly affected by the surface interaction.

We assume that the size of the coherent initial wave packet, at a distance $%
Z_{o}$ from the surface, is governed by the collimation of the incident beam
as given by the Van Cittert-Zernike theorem \cite{BornWolf}. By considering
a rectangular collimating aperture placed at a long distance $L$ from the
surface, the coherence size of the incident beam on the $Z_{o}$-plane, which
is located parallel to the surface at a $Z_{o}$ distance from it, is defined
by the complex grade of coherence, $\mu (X_{o},Y_{o})$. It reads \cite%
{BornWolf}

\begin{equation}
\left\vert \mu (X_{o},Y_{o})\right\vert ^{2}=j_{0}^{2}(\frac{\pi d_{x}}{%
\lambda _{\bot }L^{\prime }}X_{o})j_{0}^{2}(\frac{\pi d_{y}}{\lambda
L^{\prime }}Y_{o}),  \label{ucoh}
\end{equation}%
where $j_{0}(x)$ is the spherical Bessel function and $d_{x}$ and $d_{y}$
denote the lengths of the sides of the rectangular aperture, which form
angles $\theta _{x}=\pi /2-\theta _{i}$ and $\theta _{y}=0$, respectively,
with the surface plane, $\theta _{i}$ being the glancing incidence angle
(see Figs. 1 (a) and (b)). In Eq. (\ref{ucoh}) the de Broglie wave lengths $%
\lambda $ and $\lambda _{\bot }$ are defined as
\begin{equation}
\lambda =2\pi /K_{i}\text{,}\;\text{and }\;\lambda _{\bot }=\lambda /\sin
\theta _{i},  \label{lambda}
\end{equation}%
respectively, this last one being associated with the initial motion normal
to the surface plane, while $L^{\prime }=L-Z_{o}/\sin \theta _{i}$.\ For
most of the collision systems, the $Z_{o}$ value can be chosen as equal to
the lattice constant of the crystal, leading to $L^{\prime }\cong L$.

According Eq. (\ref{ucoh}) the spatial profile of the initial wave packet
can be approximated by a product of Gaussian functions,
\begin{equation}
G(\omega ,x)=[2/(\pi \omega ^{2})]^{1/4}\exp (-x^{2}/\omega ^{2}),
\label{gauss}
\end{equation}
as follows:
\begin{equation}
f_{i}(\overrightarrow{R}_{os})=G(\sigma _{x},X_{o})G(\sigma _{y},Y_{o}),
\label{ffi}
\end{equation}%
where the parameters $\sigma _{x}$ and \ $\sigma _{y}$ were derived by
fitting the complex grade of coherence, i.e. $\left\vert \mu
(X_{o},Y_{o})\right\vert ^{2}\simeq \left\vert f_{i}(\overrightarrow{R}%
_{os})\right\vert ^{2}$, reading

\begin{equation}
\sigma _{x}=\frac{\lambda _{\bot }}{\sqrt{2}}\frac{L}{d_{x}},\;\;\sigma _{y}=%
\frac{\lambda }{\sqrt{2}}\frac{L}{d_{y}}.  \label{sigmax}
\end{equation}%
The lengths $\sigma _{x}$ and $\sigma _{y}$ represent the effective widths
of the $\left\vert G(\sigma _{x},X_{o})\right\vert ^{2}$ and $\left\vert
G(\sigma _{y},Y_{o})\right\vert ^{2}$ distributions, respectively, being
defined as the corresponding root-mean-square deviations \cite{Cohen}.
Notice that these widths are associated with the \textit{transversal}
\textit{coherence} \textit{size} of the initial wave packet, magnitude that
is crucial in matter-wave interferometry \cite%
{Tonomura86,Keller00,Barrachina15}.

On the other hand, concerning the momentum profile of the initial wave
packet, as we are dealing with an incident beam with a well defined energy,
i.e. $\Delta E/E\ll 1$ \cite{Winter15}, the \textit{longitudinal coherence
length} does not play any role \cite{Tonomura86}. Consequently, the starting
momentum $\overrightarrow{K}_{o}$ satisfies the energy conservation, with $%
K_{0}=\left\vert \vec{K}_{0}\right\vert =\sqrt{2m_{P}E}$, and the
integration on \ $\overrightarrow{K}_{0}$ can be solved by making use of the
change of variables $\overrightarrow{K}_{o}=K_{o}(\cos \theta _{o}\cos
\varphi _{o},\cos \theta _{o}\sin \varphi _{o},-\sin \theta _{o})$, with $%
\theta _{o}$ and $\varphi _{o}$ varying around the incidence direction. The
shape of the corresponding angular wave packet is described again in terms
of Gaussian functions, reading
\begin{equation}
g_{i}(\overrightarrow{K}_{o})\simeq g_{i}(\Omega _{o})=G(\sigma _{\theta
},\theta _{o})G(\sigma _{\varphi },\varphi _{o}),  \label{ggi}
\end{equation}%
where $\Omega _{o}\equiv (\theta _{o},\varphi _{o})$ is de solid angle
corresponding to the $\overrightarrow{K}_{o}$ direction and the angular
widths of the $\theta _{o}$- and $\varphi _{o}$- distributions were derived
from the uncertainty principle as \cite{Cohen}
\begin{equation}
\sigma _{\theta }=\frac{\lambda _{\bot }}{2\sigma _{x}}\text{, and}\;\sigma
_{\varphi }=\frac{\lambda }{2\sigma _{y}},  \label{sigmatita}
\end{equation}%
respectively.

Replacing Eqs. (\ref{ffi}) and (\ref{ggi}) in Eq. (\ref{Aif-sivr}),\ the
extended version of the SIVR\ transition amplitude, including explicitly the
proper shape of the incident wave packet, is expressed as

\begin{eqnarray}
A_{if}^{{\small (SIVR)}} &=&\ \frac{\alpha }{\mathcal{S}}\int\limits_{%
\mathcal{S}}d\overrightarrow{R}_{os}\ f_{i}(\overrightarrow{R}_{os})\int
d\Omega _{o}\ g_{i}(\Omega _{o})  \notag \\
&&\times a_{if}^{{\small (SIVR)}}(\overrightarrow{R}_{o},\overrightarrow{K}%
_{o}),  \label{Aif-sivrn}
\end{eqnarray}%
where $a_{if}^{{\small (SIVR)}}(\overrightarrow{R}_{o},\overrightarrow{K}%
_{o})$ is given by Eq. (\ref{aif}) and $\alpha =m_{P}K_{i}$.

\section{Results}

We apply the extended SIVR method to $^{4}$He atoms elastically scattered
from a LiF(001) surface under axial surface channeling conditions since for
this collision system, diffraction patterns for different widths of the
collimating slit were reported in Ref. \cite{Winter15}. The SIVR transition
amplitude was obtained from Eq. (\ref{Aif-sivrn}) \ by employing the
MonteCarlo technique to evaluate the $\overrightarrow{R}_{os}$ and $\Omega
_{o}$ integrals, considering more than $4\times 10^{5}$ points in such an
integration. For every starting point, the partial transition amplitude $%
a_{if}^{{\small (SIVR)}}(\overrightarrow{R}_{o},\overrightarrow{K}_{o})$ was
evaluated numerically from Eq. (\ref{aif}) by employing a potential $V_{SP}$
derived from a pairwise additive hypothesis. The potential model used in
this work is the same as the one employed in Ref. \cite{Gravielle14}. It
describes the surface-projectile interaction as the sum of the static and
polarization contributions, the first of them evaluated incorporating no
local terms of the electronic density in the kinetic and exchange
potentials. The potential $V_{SP}$ also takes into account a surface
rumpling, with a displacement distance extracted from Ref. \cite%
{SchullerGrav09}. Details of the surface potential will be published
elsewhere \cite{Miragliatobe}.

In this work we vary the size of the collimating aperture keeping a fixed
incidence condition, given by helium projectiles impinging along the $%
\left\langle 110\right\rangle $ channel with a total energy $E=1$ keV and an
incidence angle $\theta _{i}=0.99$ deg. In all the cases, the distance
between the collimating aperture and the surface is chosen as $L=25$ cm, in
agreement with the experimental setup of Ref. \cite{Winter15}.

In Figs. 2 and 3 we show two-dimensional projectile distributions, as a
function of $\theta _{f}$ and $\varphi _{f}$, derived within the SIVR
approximation by considering collimation slits with the same length - $%
d_{x}=1.5$ mm - but two different widths: $d_{y}=0.2$ mm and \ $d_{y}=1.0$
mm, respectively. \ Both SIVR distributions\ reproduce quite well the
corresponding experimental ones \cite{Winter15}, also displayed in the
figures. They present the usual banana shape, characteristic of the axial
surface scattering \cite{Meyer95}, with final dispersion angles lying on a
thick annulus, whose mean radius is approximately equal to $\theta _{i}$. \
From the comparison of Figs. 2 and 3 it is clearly observed that the width
of the collimation slit strongly affects the diffraction patterns, making
the well-defined peaks present in the distributions of Fig. 2, for the more
narrow slit, completely disappear when the width of the slit is increased,
as it happens in Fig. 3. In the experimental and theoretical intensity
distributions of Fig. 3, only maxima at the rainbow deflection angles $\pm
\Theta _{rb}$ are visible. As discussed in Ref. \cite{Winter15}, this
behavior is related to the area $\mathcal{S}$ of the surface plane that is
coherently lighted by the incident beam and it will be studied in detail
within the SIVR approach.

In Eq. (\ref{Aif-sivrn}), by splitting the $\overrightarrow{R}_{os}$%
-integral on the area $\mathcal{S}$ \ into a collection of integrals over
different reduced unit cells, it is possible to express $A_{if}^{{\small %
(SIVR)}}$ \ as a product of two factors \cite{Gravielle14}:

\begin{equation}
A_{if}^{{\small (SIVR)}}\simeq A_{if,1}^{{\small (SIVR)}}\ \times F_{B},
\label{A-factors}
\end{equation}%
each of them associated with a different interference mechanism. The factor $%
A_{if,1}^{{\small (SIVR)}}$, named unit-cell form factor, is derived from
Eq. (\ref{Aif-sivrn}) by evaluating the $\overrightarrow{R}_{os}$-integral
over only one reduced unit cell, being related to supernumerary rainbows
\cite{Schuller08}. While the factor $F_{B}$ is a crystallographic factor
associated with the Bragg diffraction, which originates from the
interference of identical trajectories whose initial positions $%
\overrightarrow{R}_{os}$ are separated by a distance equal to the spacial
periodicity of the lattice. The factor $F_{B}$ depends on $%
\overrightarrow{Q}$ and the area $\mathcal{S}$\ coherently illuminated by\
the particle beam, being insensible to the potential model.

In Eq. (\ref{Aif-sivrn}) the effective area $\mathcal{S}$ coherently lighted
by the incident beam results to be $\mathcal{S\simeq D}_{x}\times \mathcal{D}%
_{y}$, where the distances $\mathcal{D}_{j}=2\sqrt{2}\sigma _{j}$ with $j=x,y
$ were determined from the $(X_{o},Y_{o})$ values for which the function $%
\left\vert \mu (X_{o},Y_{o})\right\vert ^{2}$, given by Eq. (\ref{ucoh}),
vanishes. Under typical incidence conditions for FAD, the dependence of $%
F_{B}$ on the azimuthal angle $\varphi _{f}$ becomes completely governed by
the number $n_{y}$ of reduced unit cells in the direction transversal to the
incidence channel that are illuminated by the initial wave packet, i.e. $%
n_{y}\simeq \mathcal{D}_{y}/a_{y}$ , where $a_{y}$\ is the length of the
reduced unit cell along the $\widehat{y}$ direction. For $n_{y}\gtrsim 2$
the factor $F_{B}$ gives rise to Bragg peaks placed at azimuthal angles that
verify the relation $\sin \varphi _{f}=m\lambda /a_{y}$, with $m$ an
integer, as observed in Fig. 2 where $n_{y}\simeq 4$. The relative
intensities of theses Bragg peaks are modulated by $A_{if,1}^{{\small (SIVR)}%
}$, which acts as an envelope function that can reduce or even suppress the
contribution of a given Bragg order, while the peak width is determined by $%
n_{y}$, narrowing as $n_{y}$ increases. But when the coherently illuminated
region shrinks to cover around a reduced unit cell in the transversal
direction, only the unit-cell factor is present in Eq. (\ref{A-factors}).
Consequently, the angular distribution shows structures associated with\
supernumerary rainbow maxima exclusively, as it happens in Fig. 3 where $%
n_{y}\lesssim 1$.

With the aim of studying more deeply the variation of the diffraction
patterns with the width of the slit, in Fig. 4 we display the differential
probability $dP^{{\small (SIVR)}}/d\varphi _{f}$, as a function of the
azimuthal angle $\varphi _{f}$, for different values of $d_{y}$. As given by
Eq. (\ref{sigmax}), when $d_{y}$ augments, the number $n_{y}$ of the
coherently illuminated cells decreases while the width of the Bragg peaks
increases, as observed in Fig. 4 for $d_{y}\lesssim 0.4$ mm. For wider
collimating slits Bragg peaks start to blur out, disappearing completely for
$d_{y}=0.8$ mm, where $n_{y}\simeq 1$. Therefore, varying $d_{y}$ we can
inspect two different zoologies: Bragg peaks at small $d_{y}$ values and
supernumerary rainbow peaks at large $d_{y}$.

We also analyze the influence of the length of the collimating aperture, $%
d_{x}$, on FAD patterns. In Fig. 5 we display angular projectile
distributions derived from the SIVR approach by considering a collimating
slit with the same width, $d_{y}=0.2$ mm, and three different lengths: $%
d_{x}=0.2$ , $2.0$ and \ $4.0$ mm. For a small square aperture (Fig. 5 (a)),
Bragg peaks are observed like circular spots lying on a thin ring whose
radius is equal to $\theta _{i}$, corresponding to an almost ideal elastic
rebound $\vec{K}_{i}\rightarrow \vec{K}_{f}$. But when the length of the
collimating aperture augments up to $d_{x}=$ \ $2.0$ mm (Fig. 5 (b)),
transforming the square orifice into a slit, Bragg peaks become visible like
elongated strips which are placed at slightly different radius. This effect
is even more evident in Fig. 5 (c) for $d_{x}=$ \ $4.0$ mm, where the
projectile distribution resembles the diffraction charts for different
normal energies $E_{\bot }=E\sin ^{2}\theta _{i}$. The explanation is
simple: from Eqs. (\ref{sigmax}) and (\ref{sigmatita}), if $d_{x}$ is large $%
\sigma _{\theta }$ is also large, enabling a wide spread of the impact
momentum normal to the surface plane, $\left\vert K_{oz}\right\vert
=K_{o}\sin \theta _{o}$. Such a $K_{oz}$- dispersion gives rise to the
structures along the vertical axis of Fig. 5 (c). Hence, the intensity
oscillations along the $\theta _{f}$- axis observed for long collimating
slits are probing the surface potential for different distances to the
topmost atomic plane. They might be a useful tool to explore different
distances to the surface without varying the mean value of normal energy $%
E_{\bot }$. Additionally, notice that the transversal coherence length $%
\sigma _{x}$ ($\sigma _{y}$) depends on the ratio $L/d_{x}$ ($L/d_{y}$), as
given by Eq. (\ref{sigmax}), so that any change of the collimating
conditions that kept this ratio as a constant will produce the same
interference patterns.

\section{Conclusions}

We have derived an extended version of the SIVR\ approximation \cite%
{Gravielle14} that incorporates a realistic description of the coherent
initial wave function in terms of the collimating parameters of the incident
beam. The model was applied to helium atoms grazingly impinging on a
LiF(001) surface considering a rectangular collimating aperture with
different sizes. As it was experimentally found \cite{Winter15}, the SIVR
interference patterns are strongly affected by the width of the collimating
slit, which determines the transversal length of the surface area that is
coherently illuminated by the incident wake packet. The number of lighted
reduced unit cells in the direction transversal to the incidence channel
determines the azimuthal width of the Bragg peaks, making either Bragg peaks
or supernumerary rainbows were visible.

On the other hand, the length of the collimating slit affects the polar $%
\theta _{f}$- distribution of scattered projectiles, this effect being
related to the dispersion of the component of the initial momentum
perpendicular to the surface. As the length of the collimating aperture
increases, diffraction maxima are transformed from circular spots into
elongated strips, where interference structures along the $\theta _{f}$-
axis arise for the longer slits. These findings suggest that collimating
slits with several millimeters of length might be used to probe the
projectile-surface interaction for different normal distances.

\begin{acknowledgments}
The authors acknowledge financial support from CONICET, UBA, and ANPCyT of
Argentina.
\end{acknowledgments}

\begin{figure}[tbp]
\includegraphics[width=0.4\textwidth]{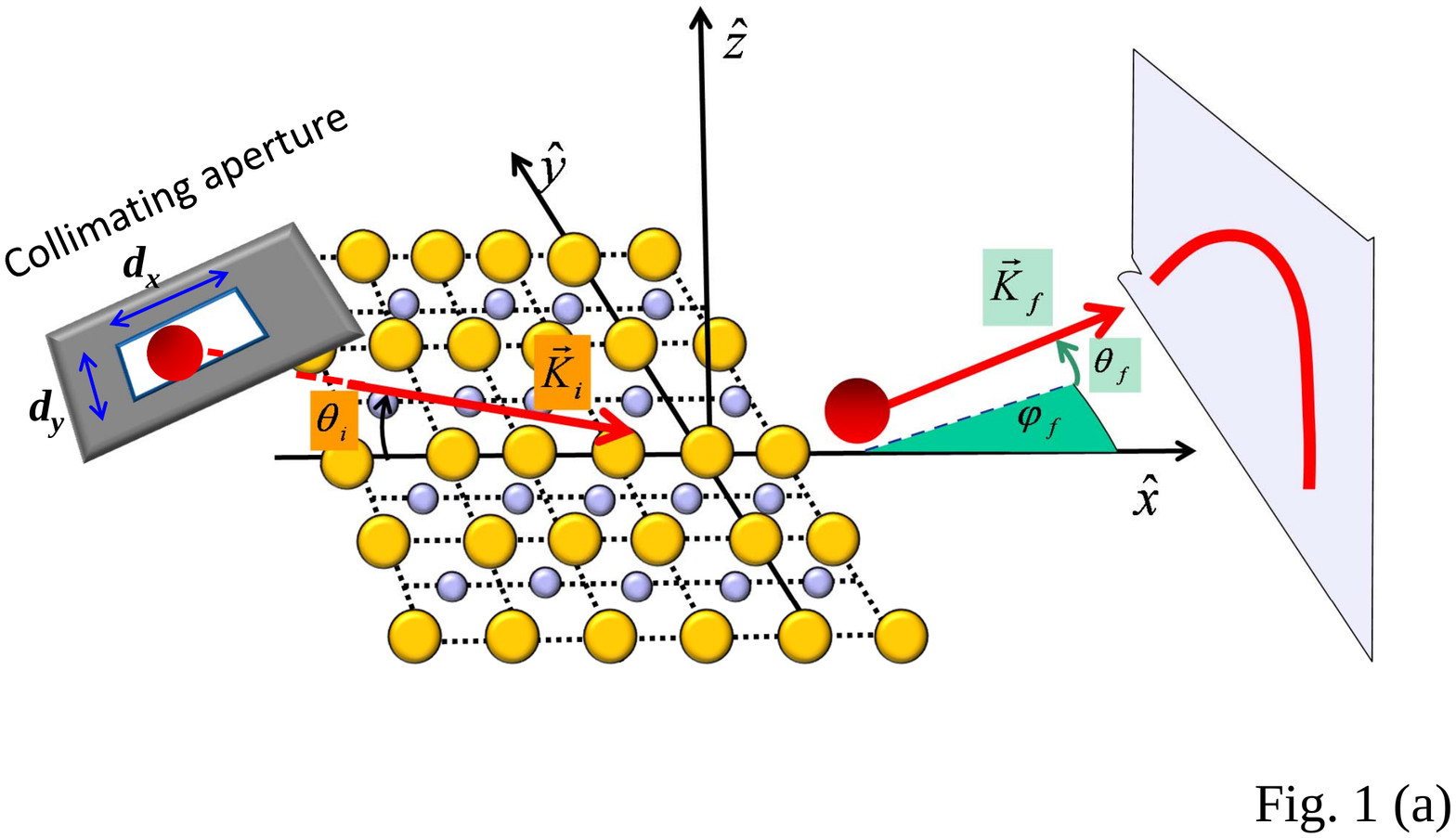} \includegraphics[width=0.4%
\textwidth]{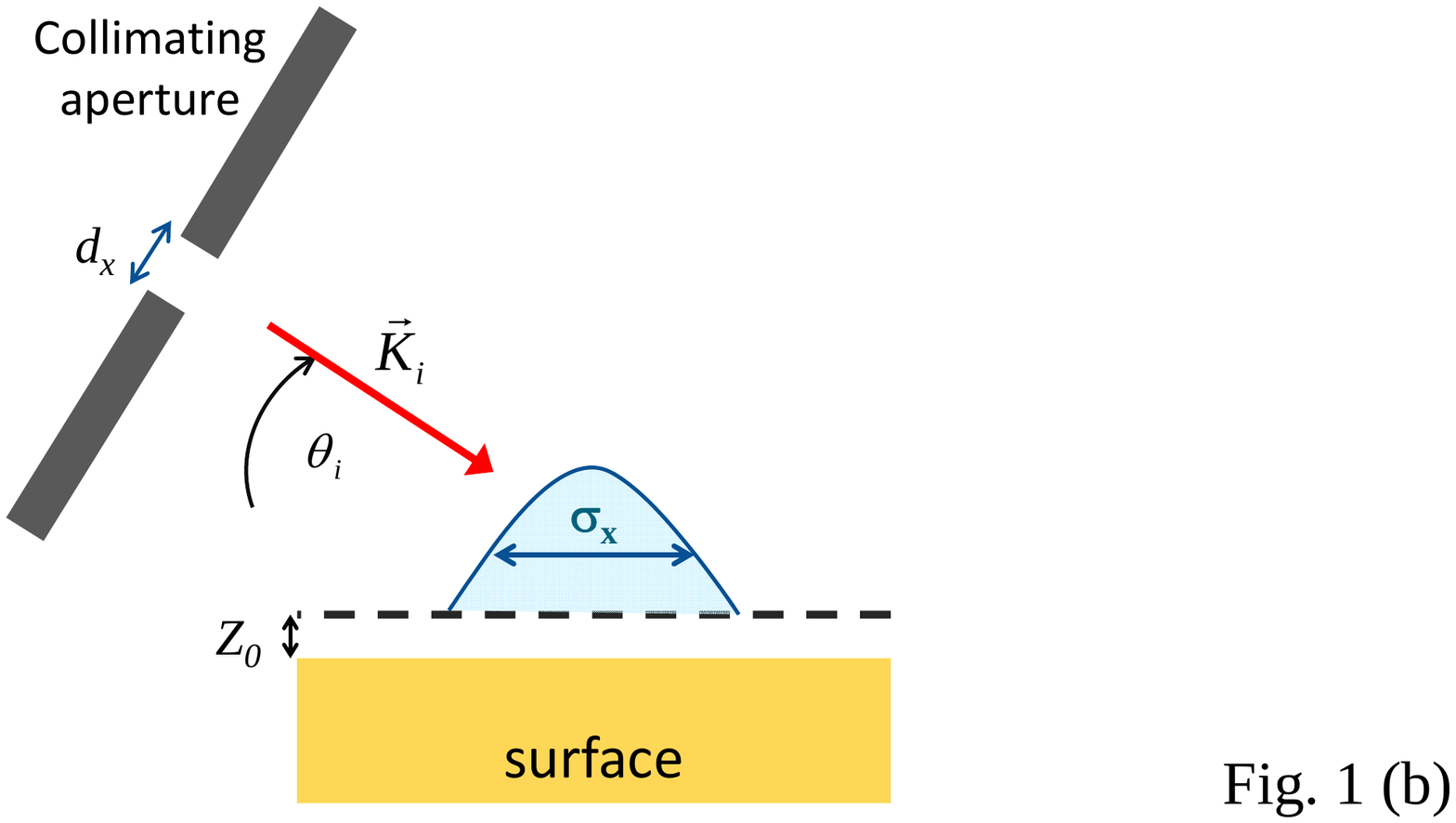}
\caption{(Color online) (a) Sketch of the FAD process, including the
collimating aperture. (b) Lateral sight of the scattering process. }
\label{Fig1}
\end{figure}

\begin{figure}[tbp]
\includegraphics[width=0.4\textwidth]{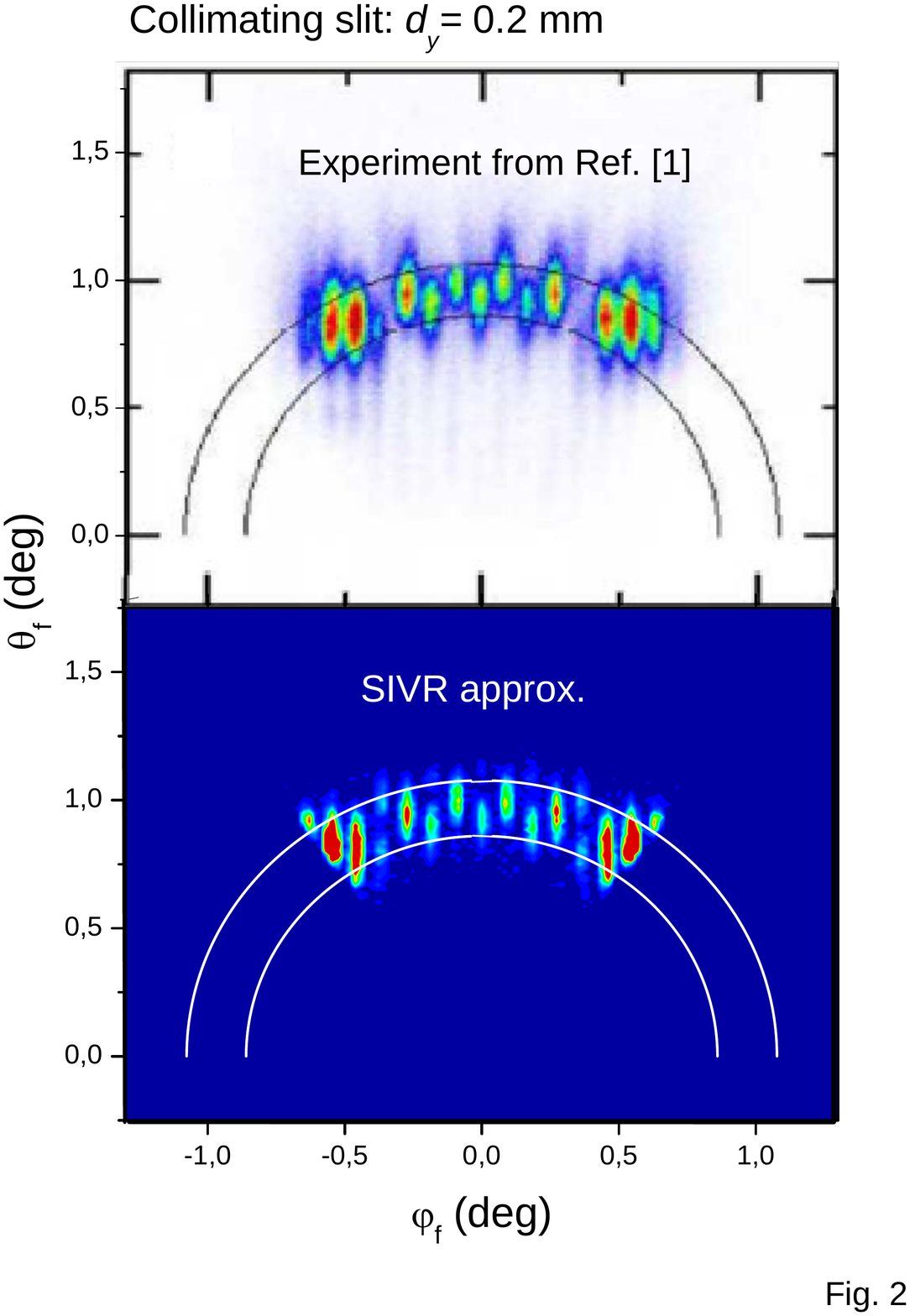}
\caption{(Color online) Two-dimensional projectile distribution, as a
function of the final dispersion angles $\protect\theta _{f}$ and $\protect%
\varphi _{f}$ , for $1$ keV $^{4}$He atoms impinging on LiF(001) along the $%
\left\langle 110\right\rangle $ direction with the incidence angle $\protect%
\theta _{i}=0.99$ deg. The incident helium beam is collimated with a
rectangular aperture of sides $d_{x}=$ \ $1.5$ mm and $d_{y}=$ \ $0.2$ mm.
Upper panel, experimental distribution extracted from Ref. \protect\cite%
{Winter15}; lower panel, SIVR distribution. }
\label{Fig2}
\end{figure}

\begin{figure}[tbp]
\includegraphics[width=0.4\textwidth]{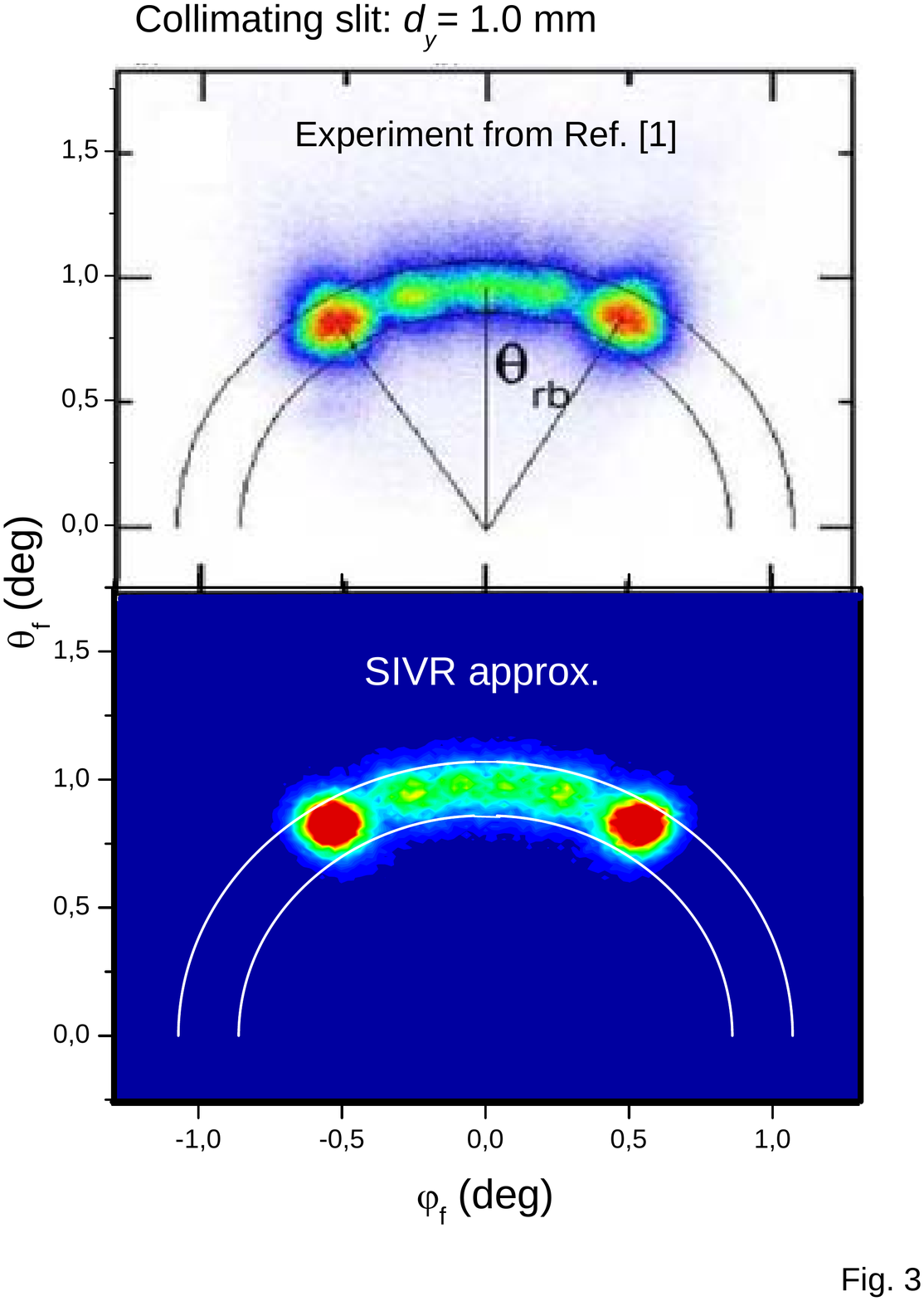}
\caption{(Color online) Similar to Fig. 2 for a collimating aperture of
sides $d_{x}=$ \ $1.5$ mm and $d_{y}=$ \ $1.0$ mm. The radial lines in the
upper panel indicate the positions of the rainbow deflection angles $\pm
\Theta _{rb}$.}
\label{Fig3}
\end{figure}

\begin{figure}[tbp]
\includegraphics[width=0.4\textwidth]{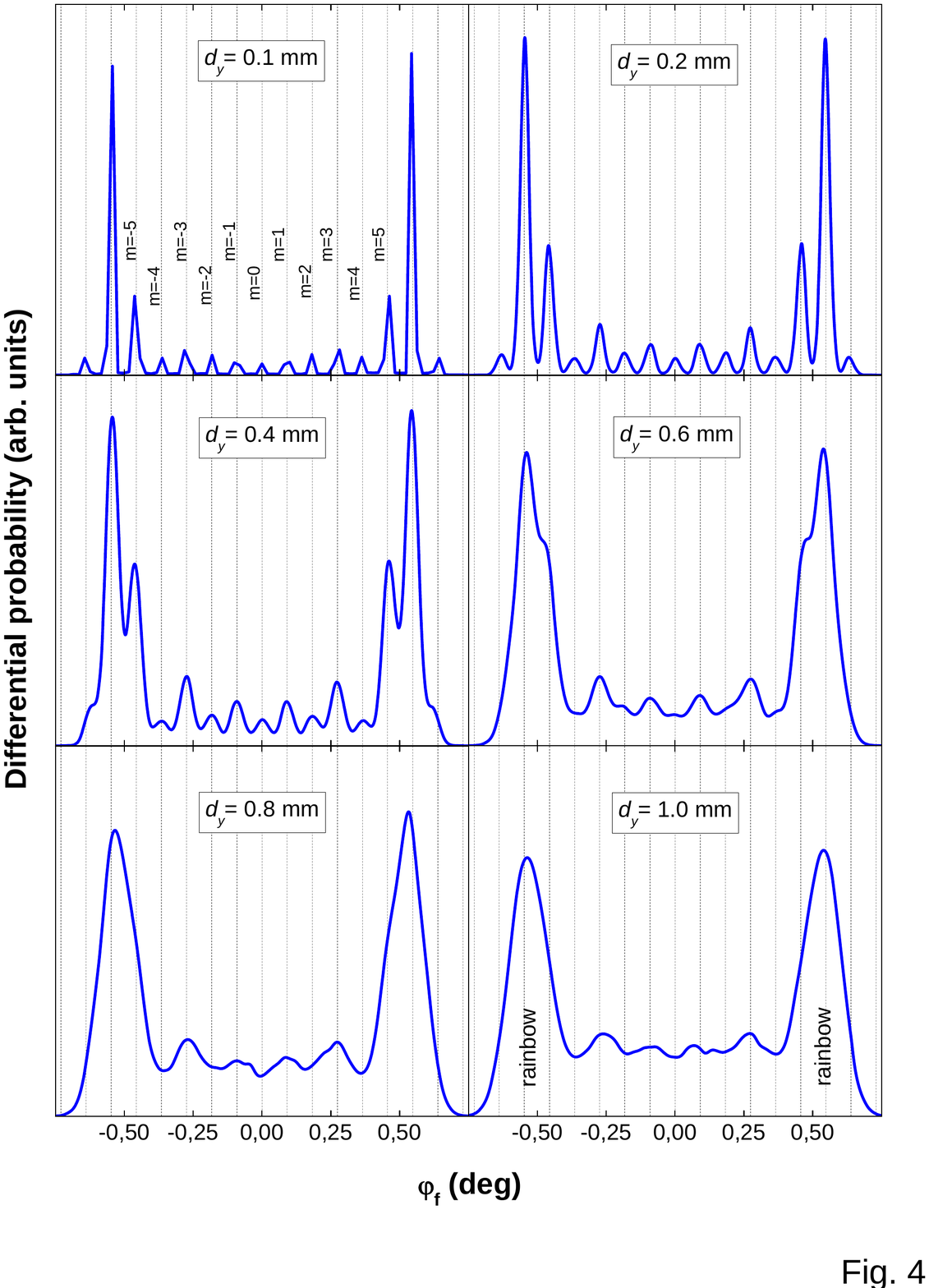}
\caption{(Color online) Azimuthal angular distribution, as a function of $%
\protect\varphi _{f}$, for $1$ keV $^{4}$He atoms impinging on LiF(001)
along the $\left\langle 110\right\rangle $ direction with the incidence
angle $\protect\theta _{i}=0.99$ deg. The incident helium beam is collimated
with a rectangular aperture of length $d_{x}=$ \ $1.5$ mm and different
widths: $\ d_{y}=$ \ $0.1$, $0.2$, $0.4$, $0.6$, $0.8$ and $1.0$ mm,
respectively. Vertical lines indicate the angular positions of Bragg peaks,
as explained in the text.}
\label{Fig4}
\end{figure}

\begin{figure}[tbp]
\includegraphics[width=0.4\textwidth]{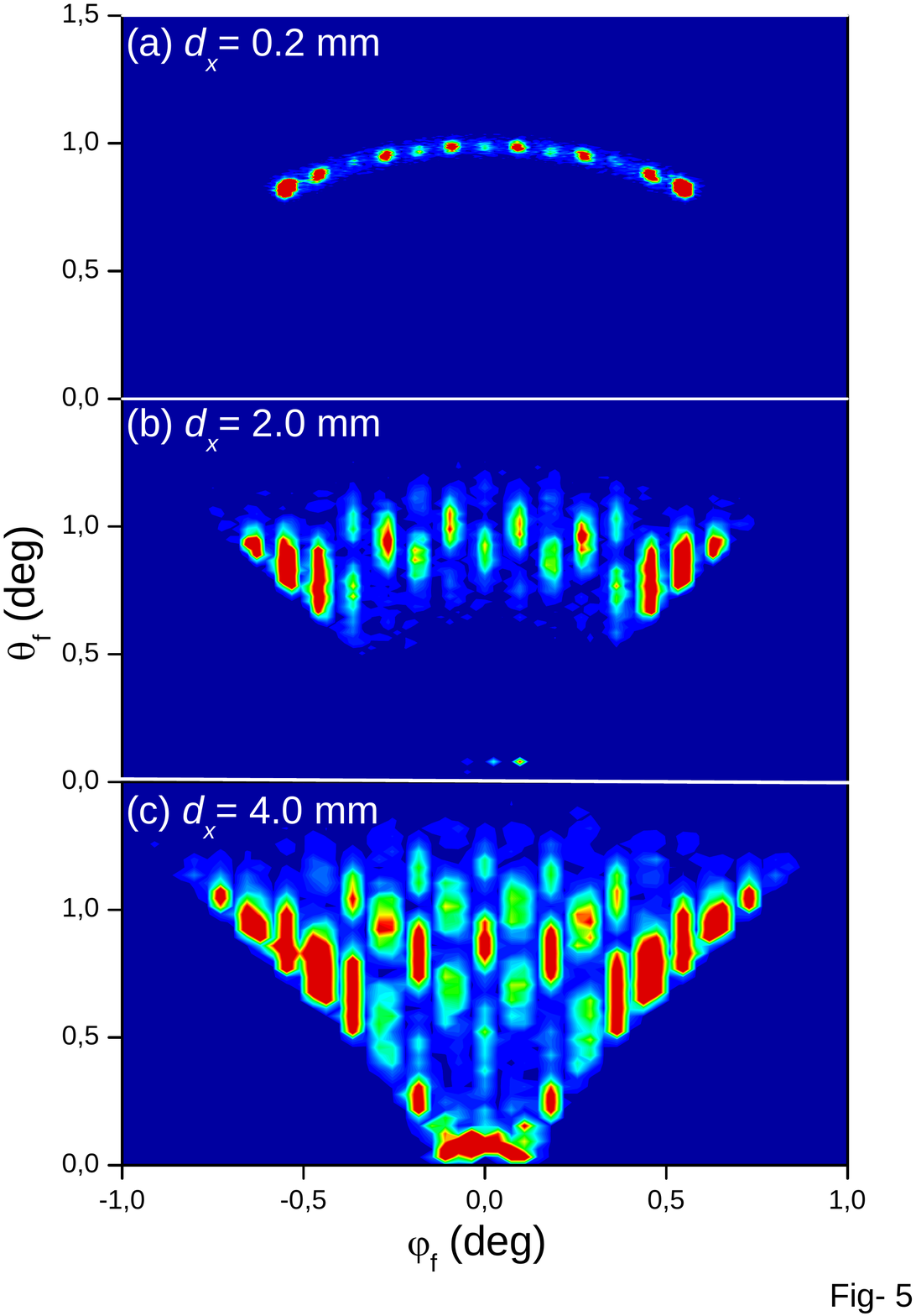}
\caption{(Color online) Similar to Fig. 2 for a collimating slit of width $%
d_{y}=$ \ $0.2$ mm and different lengths: (a) $d_{x}=$ \ $0.2$ mm, (b) $%
d_{x}=$ \ $2.0$ mm, and (c) $d_{x}=$ \ $4.0$ mm.}
\label{Fig5}
\end{figure}

\end{document}